\documentclass[twocolumn,preprintnumbers,pra]{revtex4}
\usepackage{amsmath}
\usepackage{dcolumn}
\usepackage{bm}
\usepackage{graphicx}
\usepackage{amsfonts}
\usepackage{amssymb}
\usepackage{bbm}
\usepackage{float}
\usepackage[dvipdfm]{hyperref}

\begin{document}

\title{Generalized speed and cost rate in transitionless quantum driving}
\author{Zhen-Yu Xu$^{1}$}
\email{zhenyuxu@suda.edu.cn}
\author{Wen-Long You$^{1}$}
\author{Yu-Li Dong$^{1}$}
\author{Chengjie Zhang$^{1}$}
\author{W. L. Yang$^{2}$}
\affiliation{$^{1}$College of Physics, Optoelectronics and Energy, Soochow University,
Suzhou 215006, China}
\affiliation{$^{2}$State Key Laboratory of Magnetic Resonance and Atomic and Molecular
Physics, Wuhan Institute of Physics and Mathematics, Chinese Academy of
Sciences, Wuhan 430071, China}

\begin{abstract}
Transitionless quantum driving, also known as counterdiabatic driving, is a
unique shortcut technique to adiabaticity, enabling a fast-forward evolution
to the same target quantum states as those in the adiabatic case. However,
as nothing is free, the fast evolution is obtained at the cost of stronger
driving fields. Here, given the system initially get prepared in equilibrium
states, we construct relations between the dynamical evolution speed and the
cost rate of transitionless quantum driving in two scenarios: one that
preserves the transitionless evolution for a single energy eigenstate
(individual driving), and the other that maintains all energy eigenstates
evolving transitionlessly (collective driving). Remarkably, we find that
individual driving may cost as much as collective driving, in contrast to
the common belief that individual driving is more economical than collective
driving in multilevel systems. We then present a potentially practical
proposal to demonstrate the above phenomena in a three-level Landau-Zener
model using the electronic spin system of a single nitrogen-vacancy center
in diamond.
\end{abstract}

\maketitle

\section{Introduction}

The interpretation of the energy-time uncertainty relation experienced a
long period of friendly debate after the birth of quantum mechanics. It has
now widely accepted that this uncertainty relation is not a statement about
simultaneous events but rather about the intrinsic time scale of a quantum
system in a given process to evolve to a target state \cite{MT,ML}. The
minimum period of time for the dynamical process is usually specified as the
quantum speed limit time (see recent reviews in Refs. \cite%
{QSL-review0,QSL-Review}). Since any realistic quantum system is subject to
environmental noise \cite{Book Open}, the research on the energy-time
uncertainty relation has recently been extended to general open systems \cite%
{QSL-review0,QSL-Review,QSLopen1,QSLopen2,QSLopen3,QSLopen4,QSLopen5}. This
uncertainty relation between energy and time has a large variety of
applications in quantum physics, such as exploration of the ultimate speed
of quantum computers \cite{QC-1,QC-2}, the mechanism for quantum dynamical
speedup \cite%
{speedup1,speedup2,speedup3,Xu1,speedup-exp,Xu2,speedup4,speedup5,speedup6},
the ultimate bound for parameter estimation in quantum metrology \cite%
{metrology1,metrology2}, and the efficiency of charging power in quantum
batteries \cite{QB}.

In particular, the energy-time uncertainty relation plays a key role in
understanding the principle of counterdiabatic driving \cite{Rice},\ or
transitionless quantum driving \cite{Berry}, in the realization of shortcuts
to adiabaticity \cite{STA-review}. This nonadiabatic (fast) protocol, which
reproduces the same target state as an adiabatic (slow) process, enjoys wide
applications in quantum computation and quantum thermodynamics \cite%
{STA-review,STA10,STA11,STA-exp1,STA21,STA31,shortcuts-NV,STA32,STA40,STA41,santos,STA50,STA60,song1,cost,STA61,NX,speedcost,work,santos2,song2,STA70}%
. However, there is no such thing as free speedup, e.g., the superadiabatic
route to the implementation of universal quantum computation is founded to
be bounded by the quantum speed limit \cite{santos}. Recently, a rigorous
relation between the speed of quantum evolution ($v$) and the cost rate of
the driving field ($\partial _{t}C$) has been constructed \cite{speedcost},
indicating that instantaneous manipulation is impossible as it requires an
infinite cost rate. In fact, this relation is roughly demonstrated by the
energy-time uncertainty relation ($\Delta E\sim \hbar /\tau $) since the
variance of energy $\Delta E$ is related to the cost rate of the driving
field and $\hbar /\tau \propto v$. However, this relation is obtained in a
particular case, where the driving is restricted to a particular eigenstate
(individual driving). Then questions naturally arise: Will a similar
relation still exist in a more general case where all energy eigenstates are
simultaneously driven in a transitionless way (collective driving)? Does the
cost rate of individual driving always take less consumption than the
collective case? What is the relationship of the dynamical speed between
individual and collective driving? The answer to these questions are of
great importance and may provide deeper insight into the mechanisms of
counterdiabatic driving for complex quantum systems \cite%
{STA-review,STA21,STA32,STA40}.

In this work, given the system initially get prepared in equilibrium states,
i.e., $\partial _{t}\rho |_{t=0}=0$ \cite{book-quantum}, we construct
general and concise relations (or inequality) for the speed of quantum
evolution and the cost rate both in collective and individual driving cases.
As a by-product, we find that the cost rate of individual driving may become
as large as that of collective driving, which is contrary to the common
belief that collective driving costs more than individual driving in
multilevel systems. As an example, we analyze the presented theory with a
three-level Landau-Zener (LZ) model\ and design a protocol to test the above
phenomena in an electronic spin of a single nitrogen-vacancy (NV) center in
diamond under current experimental conditions.

The rest of this paper is organized as follows. In Section II, we present
the generalized cost-cost (Sec. IIA), speed-speed (Sec. IIB), and speed-cost
(Sec. IIC) relations under individual/collective counterdiabatic driving
cases. Section III is dedicated to the three-level LZ model with the theory
introduced in Section II. An experimental analysis is performed in Section
IV with a NV center in diamond. Finally, we close with a discussion and
summary in Section V.

\section{Generalized speed and cost rate during transitionless driving}

Consider a time-dependent Hamiltonian H$(t)$ with instantaneous eigenstates
\{$\left\vert n_{t}\right\rangle $\}\ and eigenvalues \{$E_{n}(t)$\}. When $%
H(t)$ varies sufficiently slowly, the dynamics for the $n$th eigenstate $%
\left\vert n_{t}\right\rangle $ in the adiabatic approximation is $%
\left\vert \Psi _{n}(t)\right\rangle =\exp $\{$-(i/\hbar
)\int_{0}^{t}dt^{\prime }E_{n}(t^{\prime })-\int_{0}^{t}dt^{\prime
}\left\langle n_{t^{\prime }}|\partial _{t^{\prime }}n_{t^{\prime
}}\right\rangle $\}$\left\vert n_{t}\right\rangle $. The central goal of
transitionless driving is to find an auxiliary Hamiltonian $H^{A}(t)$ such
that $\left\vert \Psi _{n}(t)\right\rangle $ becomes the exact dynamical
solution of the Schr\"{o}dinger equation $i\hbar \partial _{t}\left\vert
\Psi _{n}(t)\right\rangle =\mathcal{H}(t)\left\vert \Psi
_{n}(t)\right\rangle $, where $\mathcal{H}(t)=H(t)+H^{A}(t)$. According to
the transitionless tracking algorithm, $H^{A}(t)$ can be constructed as
follows \cite{Berry},
\begin{equation}
H^{A}(t)=i\hbar \sum_{n}\partial _{t}(\left\vert n_{t}\right\rangle
\left\langle n_{t}\right\vert )\cdot \left\vert n_{t}\right\rangle
\left\langle n_{t}\right\vert .  \label{TQD}
\end{equation}%
We remark that the driving in Eq. (\ref{TQD}) guarantees that all energy
eigenstates evolve transitionlessly. For convenience, we call Eq. (\ref{TQD}%
) collective driving. However, this strong requirement can be relaxed when
we focus on a single particular eigenstate, e.g., the $n$th eigenstate $%
\left\vert n_{t}\right\rangle $, to ensure that only this eigenstate is
driven transitionlessly. In other words, we may decouple the $n$th
eigenstate from the rest eigenstates, and the corresponding auxiliary
driving Hamiltonian is modified as \cite{Rice,cost,speedcost}
\begin{equation}
H_{n}^{A}(t)=i\hbar \left[ \partial _{t}(\left\vert n_{t}\right\rangle
\left\langle n_{t}\right\vert ),\left\vert n_{t}\right\rangle \left\langle
n_{t}\right\vert \right] ,  \label{TQD-single}
\end{equation}%
which is hereinafter called individual driving for simplicity \cite{Note}.
Here the notation $[\cdot ,\cdot ]$ denotes the commutator and the subscript
$n$ represents the individual driving for the $n$th eigenstate.

\subsection{Cost rate}

A series of cost functions for transitionless driving have recently been
introduced, among which the simplest member, ignoring the set-up constant,
possesses the following form \cite{speedcost,cost}

\begin{equation}
C_{(n)}=\int_{0}^{t}dt^{\prime }\left\Vert H_{(n)}^{A}(t^{\prime
})\right\Vert ^{\alpha },  \label{cost}
\end{equation}%
where $\left\Vert X\right\Vert =\sqrt{\text{tr}\left( X^{\dag }X\right) }$
is the Frobenius norm of operator $X$, and the superscript $\alpha $ depends
on the nature of the driving field. For simplicity, we focus on the cost
rate of transitionless driving, i.e., $\partial _{t}C_{(n)}=\left\Vert
H_{(n)}^{A}(t)\right\Vert ^{\alpha }$. Through straightforward calculations
with Eq. (\ref{cost}), we obtain the cost rate for driving all energy
eigenstates as

\begin{equation}
\partial _{t}C=\hbar ^{\alpha }\left[ \sum_{n}(\langle \partial
_{t}n_{t}|\partial _{t}n_{t}\rangle +\langle n_{t}|\partial _{t}n_{t}\rangle
^{2})\right] ^{\alpha /2}  \label{cost-global}
\end{equation}%
and for driving a particular single eigenstate as

\begin{equation}
\partial _{t}C_{n}=\hbar ^{\alpha }\left[ 2\left( \langle \partial
_{t}n_{t}|\partial _{t}n_{t}\rangle +\langle n_{t}|\partial _{t}n_{t}\rangle
^{2}\right) \right] ^{\alpha /2}.  \label{cost local}
\end{equation}

Thus, it is easy to see that the relation between the collective cost rate
and the individual cost rate is given by

\begin{equation}
\partial _{t}C=\left[ \frac{1}{2}\sum_{n}(\partial _{t}C_{n})^{2/\alpha }%
\right] ^{\alpha /2}.  \label{cost-relation}
\end{equation}%
This relation implies that the individual driving cost rate may become as
large as the collective case. For instance, we have $\partial
_{t}C_{k}=\partial _{t}C$ when the condition
\begin{equation}
\partial _{t}C_{k}=\left[ \sum_{n\neq k}\left( \partial _{t}C_{n}\right)
^{2/\alpha }\right] ^{\alpha /2}  \label{condition}
\end{equation}%
\ is satisfied \cite{Note2}. Note that for $n=2$, the above condition $%
\partial _{t}C_{1}=\partial _{t}C_{2}$ is met automatically \cite{Note},
i.e., the cost rate of individual driving is equivalent to the collective
case ($\partial _{t}C=\partial _{t}C_{1}=\partial _{t}C_{2}$) for any two
energy level systems.

\subsection{Dynamical speed from the perspective of geometry}

Characterizing the dynamical speed from the perspective of geometry is
intuitionistic. Before we introduce the geometric method to define the
dynamical speed of quantum evolution, it is beneficial to recall the
definition of the instantaneous speed $v$ of an object in three-dimensional
Euclidean space with Cartesian coordinate system $v=\partial _{t}s$, where $%
s $ is the length of the trajectory with the line element $%
ds^{2}=dx^{2}+dy^{2}+dz^{2}$. Analogous to the above definition, it is very
natural to define the \textquotedblleft speed\textquotedblright\ of quantum
evolution in the space of quantum states (parameterized by $\{\sigma ^{\mu
}\}$) as
\begin{equation}
v=\partial _{t}s=\sqrt{g_{\rho }},\text{ }  \label{speed}
\end{equation}%
where the line element $ds^{2}=g_{\mu \nu }d\sigma ^{\mu }d\sigma ^{\nu }$
and $g_{\rho }=g_{\mu \nu }\partial _{t}\sigma ^{\mu }\partial _{t}\sigma
^{\nu }$ \cite{book-dg,book-gs}. The repeated Greek indices represent
summation and $g_{\mu \nu }$ is the corresponding metric. In this paper, we
adopt the quantum Fisher information metric \cite{Note3}, which is related
to the distance between quantum states $\rho _{t}$ and $\rho _{t+dt}$
measured by fidelity: $F\left( \rho _{t},\rho _{t+dt}\right) =$tr$\left[
\sqrt{\sqrt{\rho _{t}}\rho _{t+dt}\sqrt{\rho _{t}}}\right] \simeq
1-(1/2)g_{\rho }dt^{2}$ \cite{fidelity,Caves}. By employing the spectral
decomposition of quantum state $\rho _{t}=\sum_{j}p_{j}\left\vert
j\right\rangle \left\langle j\right\vert ,$ $g_{\rho }$ can be written in an
explicit form \cite{book-gs,fidelity,Caves}
\begin{equation}
g_{\rho }=\frac{1}{4}\sum_{j}\frac{(\partial _{t}p_{j})^{2}}{p_{j}}+\frac{1}{%
2}\sum_{j\neq l}\frac{(p_{j}-p_{l})^{2}}{p_{j}+p_{l}}\left\vert \left\langle
j_{t}|\partial _{t}l_{t}\right\rangle \right\vert ^{2}.  \label{metric0}
\end{equation}%
It is easy to check that if the state is pure, e.g., $\rho _{t}=\left\vert
\Phi _{t}\right\rangle \left\langle \Phi _{t}\right\vert ,$ Eq. (\ref%
{metric0}) reduces to

\begin{equation}
g_{\Phi }=\frac{\left\vert \left\langle \partial _{t}\Phi _{\perp }|\partial
_{t}\Phi _{t}\right\rangle \right\vert ^{2}}{\left\langle \partial _{t}\Phi
_{\perp }|\partial _{t}\Phi _{\perp }\right\rangle },  \label{metric00}
\end{equation}%
where the unnormalized state $\left\vert \partial _{t}\Phi _{\perp
}\right\rangle =\left\vert \partial _{t}\Phi _{t}\right\rangle -\left\langle
\Phi _{t}|\partial _{t}\Phi _{t}\right\rangle \left\vert \Phi
_{t}\right\rangle $ is the component of $\left\vert \partial _{t}\Phi
_{t}\right\rangle $ orthogonal to $\left\vert \Phi _{t}\right\rangle $ \cite%
{Caves,Xu}.

For simplicity, we consider the initial states get prepared in equilibrium,
i.e., $\partial _{t}\rho |_{t=0}=0$ \cite{book-quantum}. Therefore, $\rho
_{0}$ and $H$ can be simultaneously diagonalized, thus providing the
possibilities to establish a link between cost rate and speed of evolution.
We first consider an initial state in the form of canonical ensemble $\rho
_{0}=\exp [-H(0)/(kT)]/Z=\sum_{n}p_{n}\left\vert n_{0}\right\rangle
\left\langle n_{0}\right\vert $, where $Z=$tr$[e^{-H(0)/(kT)}]$ is the
partition function with temperature $T$, $k$ is the Boltzmann constant, and $%
p_{n}=\exp [-E_{n}(0)/(kT)]/Z$ \cite{book-quantum}. We remark that under
transitionless driving, $p_{n}$ is time-independent, i.e., $\partial
_{t}p_{n}=0$. Therefore, Eq. (\ref{metric0}) reduces to
\begin{equation}
g_{\rho }=\frac{1}{2}\sum_{m\neq n}\frac{\left( p_{m}-p_{n}\right) ^{2}}{%
p_{m}+p_{n}}|\left\langle m_{t}|\partial _{t}n_{t}\right\rangle |^{2}.
\label{metric}
\end{equation}%
As a special case when the system is initially in the $n$th eigenstate $%
\left\vert n_{0}\right\rangle $, according to Eq. (\ref{metric00}), the
quantum Fisher information metric continuously reduces to the well-known
Fubini--Study metric as follows:
\begin{eqnarray}
g_{n} &=&\frac{\left\vert \langle \partial _{t}n_{\perp }|\partial
_{t}n_{t}\rangle \right\vert ^{2}}{\langle \partial _{t}n_{\perp }|\partial
_{t}n_{\perp }\rangle },  \notag \\
&=&\langle \partial _{t}n_{t}|\partial _{t}n_{t}\rangle +\langle
n_{t}|\partial _{t}n_{t}\rangle ^{2},  \label{metric2}
\end{eqnarray}%
where $\left\vert \partial _{t}n_{\perp }\right\rangle =\left\vert \partial
_{t}n_{t}\right\rangle -\left\langle n_{t}|\partial _{t}n_{t}\right\rangle
|n_{t}\rangle $.

Here, it is convenient to establish a relation of quantum speed between the
collective and the individual driving. According to Eqs. (\ref{speed}), (\ref%
{metric}), and (\ref{metric2}) we have%
\begin{eqnarray}
v &=&\sqrt{\frac{1}{2}\sum_{m\neq n}\frac{\left( p_{m}-p_{n}\right) ^{2}}{%
p_{m}+p_{n}}|\left\langle m_{t}|\partial _{t}n_{t}\right\rangle |^{2}},
\notag \\
&\leq &\sqrt{\frac{1}{2}\sum_{m\neq n}(p_{m}+p_{n})|\left\langle
m_{t}|\partial _{t}n_{t}\right\rangle |^{2}},  \notag \\
&=&\sqrt{\sum_{m\neq n}p_{n}|\left\langle m_{t}|\partial
_{t}n_{t}\right\rangle |^{2}},  \notag \\
&=&\sqrt{\sum_{n}p_{n}\left( \langle \partial _{t}n_{t}|\partial
_{t}n_{t}\rangle +\langle n_{t}|\partial _{t}n_{t}\rangle ^{2}\right) },
\notag \\
&=&\sqrt{\sum_{n}p_{n}(v_{n})^{2}},  \label{speedX}
\end{eqnarray}%
where $v_{n}=\sqrt{\langle \partial _{t}n_{t}|\partial _{t}n_{t}\rangle
+\langle n_{t}|\partial _{t}n_{t}\rangle ^{2}}$. Note that $\left(
p_{m}-p_{n}\right) ^{2}\leq \left( p_{m}+p_{n}\right) ^{2}$ is employed in
the second line of Eq. (\ref{speedX}), and the equality in the second line
is\ achieved when the system is initially prepared in a single eigenstate.

\subsection{Relationship between cost rate and dynamical speed}

Clearly, with Eqs. (\ref{cost local}), (\ref{speed}), and (\ref{metric2}),
the cost rate and the dynamical speed of transitionless quantum driving for
a single eigenstate $\left\vert n_{t}\right\rangle $ is obtained as%
\begin{equation}
v_{n}=\frac{\sqrt[\alpha ]{\partial _{t}C_{n}}}{\sqrt{2}\hbar }.
\label{speed-cost1}
\end{equation}%
We note that the above relation bears resemblance to the speed and the cost
rate relation first presented in Ref. \cite{speedcost} but in a more concise
form. However, the relation between the cost rate and the dynamical speed by
collective transitionless driving is not as concise as Eq. (\ref{speed-cost1}%
). For the collective transitionless driving case ($n\geq 2$), with Eqs. (%
\ref{cost-global}) and (\ref{speedX}), we have%
\begin{eqnarray}
v &<&\sqrt{\sum_{n}(\langle \partial _{t}n_{t}|\partial _{t}n_{t}\rangle
+\langle n_{t}|\partial _{t}n_{t}\rangle ^{2})}  \notag \\
&=&\frac{\sqrt[\alpha ]{\partial _{t}C}}{\hbar }.  \label{speed-cost2}
\end{eqnarray}

Equations (\ref{cost-relation}), (\ref{speedX}), (\ref{speed-cost1}), and (%
\ref{speed-cost2}), which reflect the general relations between the
dynamical speed and the cost rate in shortcuts to adiabaticity for both
collective and individual transitionless driving, are the main contributions
in this paper. The properties of these relations are explored in the
following pedagogical nontrivial example.

\section{Three-level Landau-Zener tunneling model}

In this section, we consider the Landau-Zener tunneling model in the
simplest multilevel system, i.e., the three-level system, with the following
Hamiltonian
\begin{equation}
H(t)=\gamma _{e}\mathbf{B}(t)\cdot \mathbf{S},  \label{LZ}
\end{equation}%
where $\gamma _{e}$ is the electronic gyromagnetic ratio, $\mathbf{B}%
(t)=\{\Delta ,0,\lambda (t)\}/\gamma _{e}$\ is the magnetic field applied
along the $x$ and $z$ directions, 
and $\mathbf{S}=$\{$S_{x},S_{y},S_{z}$\} is the $S=1$ electron spin operator
\cite{spin}. Here 
$\Delta $ denotes the minimum energy separation and $\lambda (t)$
characterizes the strength of a controllable driving field. 

According to the method introduced in Ref. \cite{Berry}, it is convenient to
validate 
that the collective transitionless driving for Eq. (\ref{LZ}) is given by
\begin{equation}
H^{A}(t)=V(t)S_{y},  \label{LZ-A}
\end{equation}%
with $V(t)=-\Delta \partial _{t}\lambda (t)/[\Delta ^{2}+\lambda ^{2}(t)]$,
which is similar to the two-level case \cite{Berry,shortcuts-NV}. The given
control field is scanned linearly within $t\in \lbrack 0,\tau ],$ i.e., $%
\lambda (t)=\kappa (2t/\tau -1),$ where $\kappa $ is related to the strength
of the driving field. Considering the nature of the applied fields, we adopt
the parameter $\alpha =2$ \cite{cost} for evaluating the cost rate of
transitionless driving.

\begin{figure}[tbp]
\centering{}\includegraphics[width=3.4in]{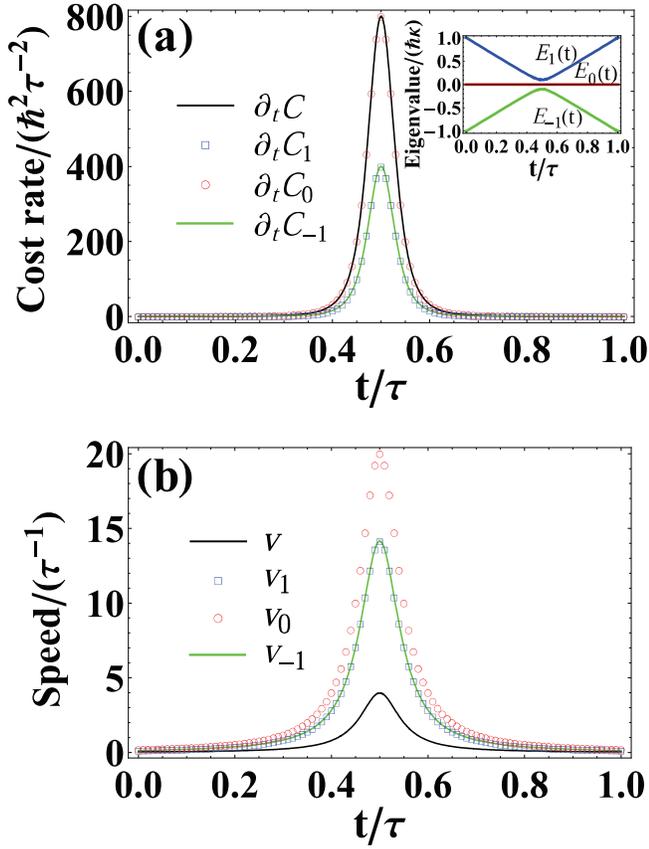}
\caption{(Color online) (a) The cost rate (divided by $\hbar ^{2}\protect%
\tau ^{-2}$) of collective ($\partial _{t}C$) and individual ($\partial
_{t}C_{1},\partial _{t}C_{0},\partial _{t}C_{-1}$) transitionless quantum
driving for the three-level LZ model evolved through the avoided crossing by
$\protect\lambda (t)=\protect\kappa (2t/\protect\tau -1)$, and $\Delta /%
\protect\kappa =0.1$. The inset shows the energy eigenvalues (divided by $%
\hbar \protect\kappa $) of the LZ model of Eq. (\protect\ref{LZ}) with the
above parameters. (b) The corresponding speed (divided by $\protect\tau %
^{-1} $) of states under collective and individual transitionless driving.
For collective driving, we have set $\hbar \protect\kappa /(kT)=1/2$ for the
initial canonical ensemble state as an example.}
\end{figure}

We first employ the above driving protocol to analyze the collective and
individual cost rates during transitionless quantum driving. The results are
shown in Fig. 1(a) with $\Delta /\kappa =0.1$ as an example. In general, the
peaks in Fig. 1(a) illustrate that more resources or higher cost rates are
required to realize transitionless driving in the neighborhood of an avoided
crossing. Obviously, the individual driving of eigenstates $E_{\pm 1}(t)$
costs less than the collective driving, which can easily be understood in
terms of Eqs. (\ref{cost-global}) and (\ref{cost local}) as $\partial
_{t}C_{+1}=\partial _{t}C_{-1}=(1/2)\partial _{t}C_{0}<\partial _{t}C$.
However, an interesting phenomenon occurs. The individual driving for
eigenstate $E_{0}(t)$ (red circle) costs as much as the collective driving
(black curve), in contrast to the common belief that the individual
transitionless driving leads to less consumption for multilevel quantum
systems. In fact, since $\partial _{t}C_{0}=2\partial _{t}C_{\pm 1}=\partial
_{t}C_{+1}+\partial _{t}C_{-1}$, which is just the condition of Eq. (\ref%
{condition}) in the three level case. Therefore, we have $\partial
_{t}C=\partial _{t}C_{0}$. Physically, this can be interpreted by means of
the configuration of eigenstates in Eq. (\ref{LZ}) shown in the insets of
Fig. 1(a). In order to achieve the individual driving of $E_{0}(t)$,
transitions from $E_{0}(t)$ to both $E_{-1}(t)$ and $E_{1}(t)$ should be
avoided. In turn, the eigenstates $E_{\pm }(t)$ will not transit to $E_{0}(t)
$, and their mutual transitions are prohibited, which is the principle of
collective driving.

\begin{figure}[t]
\centering{}\includegraphics[width=3.4in]{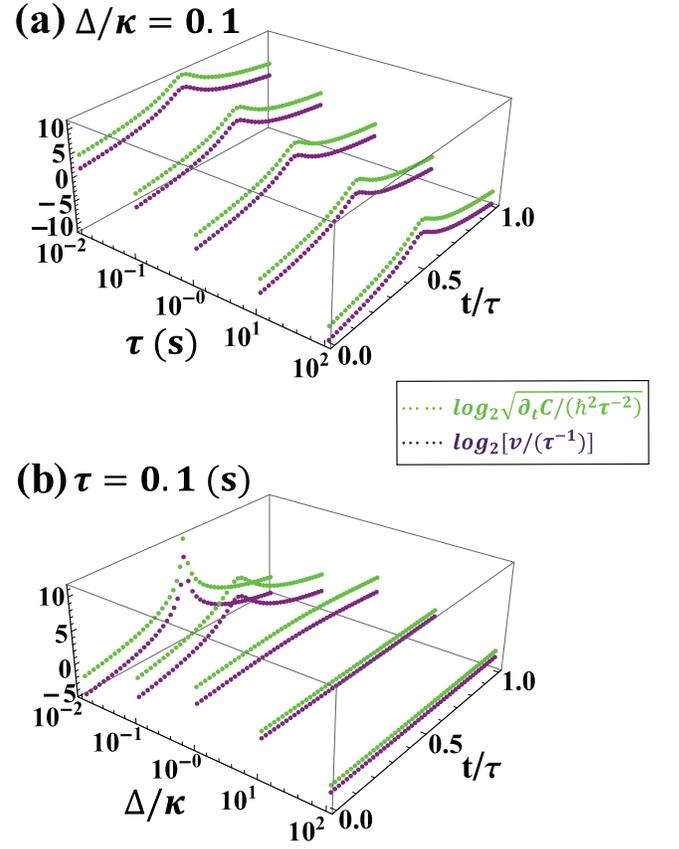}
\caption{(Color online) Time dependence of the dimensionless cost rate log$%
_{2}\protect\sqrt{\partial _{t}C/(\hbar ^{2}\protect\tau ^{-2})}$ and speed
log$_{2}[v/(\protect\tau ^{-1})]$ during collective transitionless quantum
driving for the three-level LZ model. (a) Different choices of time duration
$\protect\tau $ with fixed energy splitting $\Delta /\protect\kappa =0.1$.
(b) Various energy splitting $\Delta $ with fixed time duration $\protect%
\tau =0.1$ $(s)$. All other parameters are the same as in Fig. 1.}
\end{figure}

In addition, the tendency between the cost rate and the speed can be seen by
comparison with Fig. 1(b), where the instantaneous speed of states for the
above two scenarios is depicted. We note that although the cost rates for
the collective driving ($\partial _{t}C$) and the individual driving ($%
\partial _{t}C_{0}$) are equal in this model, the corresponding dynamical
speed does not possess such a property, i.e., $v<v_{0}$ [see in Fig. 1(b)].

We then examine the instantaneous cost rate and the speed under
transitionless driving with different driving time durations $\tau $ and
energy splittings $\Delta $. For convenience, we focus on the collective
driving, and the values are reported in the form of base-2 logarithm. As
clearly shown in Fig. 2(a), if the energy splitting is fixed, e.g., $\Delta
/\kappa =0.1$, a higher cost rate is required to rapidly go through the
avoided crossing with a faster dynamical speed to realize the transitionless
driving, which is in agreement with Eq. (\ref{speed-cost2}). When
approaching the adiabatic limit, e.g., $\tau =100$ (s), almost no
transitionless driving is needed ($\partial _{t}C\rightarrow 0$). On the
other hand, if the energy splitting $\Delta $ is sufficiently large, e.g., $%
\Delta /\kappa =100$ in Fig. 2(b), only a little cost is required to achieve
the transitionless driving.

\section{Possible experimental realization using NV centers in diamond}

\begin{figure}[tbp]
\centering{}\includegraphics[width=3.4in]{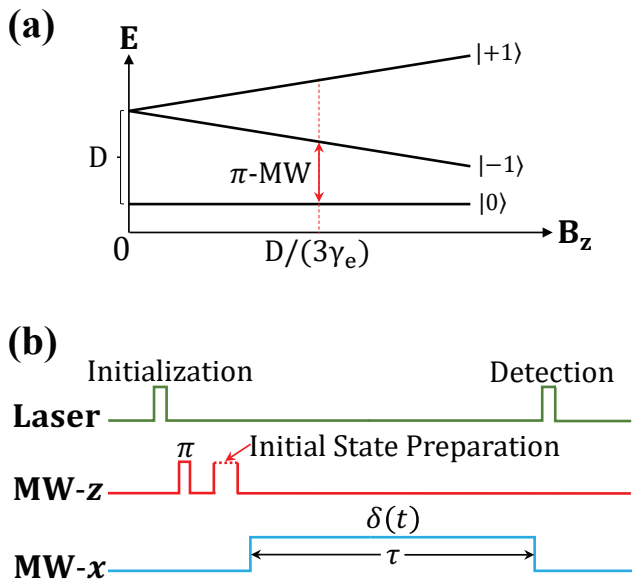}
\caption{(Color online) (a) The energy-level diagram for the electronic spin
ground triplet state of a single NV center with a zerofield splitting $D$. A
polarized $\protect\pi -$MW pulse is performed on $\left\vert 0\right\rangle
$ and $\left\vert -1\right\rangle $ at $B_{z}=D/(3\protect\gamma _{e})$ to
achieve the required Hamiltonian $H_{NV}(t)=\protect\omega _{0}S_{z}$ with $%
\protect\omega _{0}=2D/3$. (b) Diagram of the schematic experimental pulse
sequences employed to realize the three level LZ model under tranistionless
driving. The electronic spin state is initialized by 532 nm laser. The
initial equilibrium states can be prepared by a serial of MW pulses. Then a
microwave field $B_{x}(t)=\protect\delta (t)/\protect\gamma _{e}$ with $%
\protect\delta (t)=2[\Delta \cos \protect\varepsilon (t)-V(t)\sin \protect%
\varepsilon (t)]$ and $\partial _{t}\protect\varepsilon (t)=\protect\omega %
_{0}-$ $\protect\lambda (t)$, is performed along the $x$-axis of the NV
center. Tomography is then performed if we would like to detect the
instantaneous speed and cost rate ($\protect\tau $ is short enough). We may
also roughly estimate the average speed $\bar{v}$ by only record the time
duration $\protect\tau $ (relatively long) of the $B_{x}(t)$ performed on
the electronic spin, since $\protect\tau \varpropto 1/\bar{v}$, where $\bar{v%
}=(1/\protect\tau )\protect\int_{0}^{\protect\tau }dt^{\prime }v.$}
\end{figure}

The NV center spins in diamond possess long coherence time at room
temperature and high sensitivity to external signals. These properties make
the NV center a promising candidate for quantum computation \cite{NV-Review1}
and quantum sensors \cite{NV-Review2}. Here, we first outline a possible
implementation of detecting the collective cost rate of transitionless
driving and the dynamical speed for the three-level LZ model by using the
electron spin of a single NV center in diamond. The key procedures includes
the preparation of initial states, the realization of the three-level LZ
Hamiltonian under transitionless driving, i.e., Eqs. (\ref{LZ}) and (\ref%
{LZ-A}), and the detection of speed and cost rate.

We first apply a static magnetic field to the [111] axis (taken as the $z$%
-axis) of the NV center and employ the ground electronic spin state $%
\left\vert -1\right\rangle ,$ $\left\vert 0\right\rangle ,$ and $\left\vert
+1\right\rangle $ as the qutrit. According to Eq. (\ref{LZ}), it is
convenient to check that the energy eigenstates of the initial three-level
LZ Hamiltonian are just $\left\vert -1\right\rangle ,$ $\left\vert
0\right\rangle ,$ and $\left\vert +1\right\rangle $. Therefore, preparation
of an initial single energy eigenstate is achievable by microwave (MW)
pulses \cite{LZS-NV,LZS-NV2}. For an initial canonical ensemble, it can be
prepared by waiting for a certain time for the dephasing from a
superposition state \cite{ion}. On the other hand, the Hamiltonian
describing the electronic spin of the NV center takes the form \cite%
{NV-Review1,NV-Review2}%
\begin{equation}
H_{NV}(t)=DS_{z}^{2}/\hbar +\gamma _{e}B_{z}S_{z},  \label{H-NV0}
\end{equation}%
where the zero-field splitting $D=2.870$ (GHz), and the electronic
gyromagnetic ratio $\gamma _{e}=28.02$ (GHz/T). We select $B_{z}=D/(3\gamma
_{e})$ and apply a polarized $\pi -$MW pulse on $\left\vert 0\right\rangle $
and $\left\vert -1\right\rangle $ after biasing all three energy levels by $%
2\hbar D/3$. Then, Eq. (\ref{H-NV0}) reduces to $H_{NV}(t)=\omega _{0}S_{z}$
with $\omega _{0}=2D/3$ [see Fig. 3(a)]. According to Eqs. (\ref{LZ}) and (%
\ref{LZ-A}), as long as we select appropriate $\lambda (t)$ and $V(t)$, the
collective transitionless driving is immediately achievable. However, the LZ
avoided crossing cannot be realized directly by electron spin resonance
because it requires the microwave strength approaching $4D/(3\gamma
_{e})\simeq 0.137$ (T), far beyond the current experimental conditions \cite%
{LZS-NV,LZS-NV2,shortcuts-NV}. In light of the method introduced in Refs.
\cite{LZS-NV,LZS-NV2,shortcuts-NV}, our collective transitionless driving
for a three-level LZ model can also be realized in a rotating frame if we
apply a microwave field $B_{x}(t)=\delta (t)/\gamma _{e}$, where $\delta
(t)=2[\Delta \cos \varepsilon (t)-V(t)\sin \varepsilon (t)]$ and $\partial
_{t}\varepsilon (t)=\omega _{0}-$ $\lambda (t)$, along the $x$-axis of the
NV center. Thus, the total Hamiltonian in the laboratory frame is $\mathcal{H%
}(t)=\gamma _{e}B_{x}(t)S_{x}+\omega _{0}S_{z}$. By transferring to the
rotating frame with $U=$exp$[i\varepsilon (t)S_{z}/\hbar ]$, we obtain
\begin{eqnarray}
\mathcal{\tilde{H}}(t) &=&UH(t)U^{\dag }+i\hbar (\partial _{t}U)U^{\dag },
\notag \\
&=&\lambda (t)S_{z}+U\delta (t)S_{x}U^{\dag },  \notag \\
&\simeq &\Delta S_{x}+V(t)S_{y}+\lambda (t)S_{z},  \label{H-RF}
\end{eqnarray}%
where a rotating wave approximation, ignoring the fast-oscillating items exp[%
$\pm 2i\varepsilon (t)$], is employed in the deduction of the third line.
Clearly, the above equation yields the exact Hamiltonian presented in Eqs. (%
\ref{LZ}) and (\ref{LZ-A}). In this rotating frame, $\Delta $ and $\lambda
(t)$ are, respectively, controlled by the power and frequency of the
microwave field in the $x$-axis, which in turn determines the corresponding
counterdiabatic field $V(t)$. Thus, the cost rate of the transitionless
quantum driving field and the speed of evolution can be completely
controlled the microwave field $B_{x}(t)$. According to Eqs. (\ref%
{cost-global}) and (\ref{speedX}), the detection of the speed and cost rate
in experiment is flexible, which can be realized by the tomography of
instantaneous density matrix. In addition, even with no tomography, we can
also roughly estimate the average speed of quantum evolution by detecting
the duration of the evolved time $\tau $, since $\tau \varpropto 1/\bar{v}$,
where $\bar{v}=(1/\tau )\int_{0}^{\tau }dt^{\prime }v$. A diagram of the
above schematic experimental pulse sequences is depicted in Fig. 3(b).

Because the individual transitionless driving is equivalent to a
collectively driving two-level system \cite{cost,Note}; thus, the
corresponding experimental methods can refer to the two-level case in Ref.
\cite{shortcuts-NV}. Together with above analysis, the phenomenon that cost
rate of individual driving may be as large as collective driving can also be
verified with above three-level LZ model in NV centers.

\section{DISCUSSIONS AND CONCLUSIONS}

Though our present example mainly focuses on a three-level LZ model, the
cost rate and speed relations of collective driving and individual driving
presented in Sec. II are applicable to any multilevel systems. Therefore,
further study on more complicated multilevel ($n\geq 4$) physical systems
will be of great interest and importance in the field of shortcuts to
adiabaticity. On the other hand, quantum thermodynamics processes of
experimental implementation at the fundamental level of a single spin is now
emerging, e.g., a single-spin test with a single ultracold $^{40}$Ca$^{+}$
trapped ion has been employed to verify the Jarzynski-Related information
equality \cite{ion}. Therefore, in addition to NV centers, it would also be
desirable and interesting to further investigate our theory with trapped ion
systems in experiment.

In summary, general relations between the dynamical speed and the cost rate
of individual/collective transitionless driving have been constructed, which
provide a unified way to explore the cost-cost, speed-speed, and speed-cost
relations under individual/collective counterdiabatic driving. In
particular, the counterintuitive phenomenon that the cost rate of individual
driving can be as large as the corresponding collective driving in
multilevel systems has been discovered and illustrated in a three-level LZ
model. We have also proposed a possible experimental verification of this
phenomenon in the electron spin of a single NV center in diamond. We expect
these studies to contribute to the identification of the physical mechanisms
for the costs of shortcuts to adiabaticity and its experimental examination
in simple/complex quantum systems.

\section*{ACKNOWLEDGMENTS}

This work was supported by the National Natural Science Foundation of China
under Grant Nos. 11674238, 11474211, 11204196, 11504253 and 11574353.


\begin{thebibliography}{99}
\bibitem{MT} L. Mandelstam and I. Tamm, J. Phys. (USSR) \textbf{9}, 249
(1945).

\bibitem{ML} N. Margolus and L. B. Levitin, Phys. D \textbf{120}, 188(1998).

\bibitem{QSL-review0} M. R. Frey, Quantum Inf. Process. \textbf{15}, 3919
(2016).

\bibitem{QSL-Review} S. Deffner and S. Campbell, J. Phys. A, \textbf{50},
453001 (2017).

\bibitem{Book Open} H.-P. Breuer and F. Petruccione, \textit{The Theory of
Open Quantum Systems} (Oxford University Press, Oxford, 2007).

\bibitem{QSLopen1} M. M. Taddei, B. M. Escher, L. Davidovich, and R. L. de
Matos-Filho, Phys. Rev. Lett. \textbf{110}, 050402 (2013).

\bibitem{QSLopen2} A. del Campo, I. L. Egusquiza, M. B. Plenio, and S. F.
Huelga, Phys. Rev. Lett. \textbf{110}, 050403 (2013).

\bibitem{QSLopen3} S. Deffner and E. Lutz, Phys. Rev. Lett. \textbf{111},
010402 (2013).

\bibitem{QSLopen4} I. Marvian and D. A. Lidar, Phys. Rev. Lett. \textbf{115}%
, 210402 (2015).

\bibitem{QSLopen5} D. P. Pires, M. Cianciaruso, L. C. C\'{e}leri, G. Adesso,
D. O. Soares-Pinto, Phys. Rev. X \textbf{6}, 021031 (2016).

\bibitem{QC-1} S. Lloyd, Nature (London) \textbf{406}, 1047 (2000).

\bibitem{QC-2} I. L. Markov, Nature (London) \textbf{512}, 147 (2014).

\bibitem{speedup1} V. Giovannetti, S. Lloyd, and L. Maccone, Phys. Rev. A
\textbf{67}, 052109 (2003).

\bibitem{speedup2} J. Batle, M. Casas, A. Plastino, and A. R. Plastino,
Phys. Rev. A \textbf{72}, 032337 (2005).

\bibitem{speedup3} A. Borras, M. Casas, A. R. Plastino, and A. Plastino,
Phys Rev. A \textbf{74}, 022326 (2006).

\bibitem{Xu1} Z.-Y. Xu, S. Luo, W. L. Yang, C. Liu, and S. Zhu, Phys. Rev. A
\textbf{89}, 012307 (2014).

\bibitem{speedup-exp} A. D. Cimmarusti, Z. Yan, B. D. Patterson, L. P.
Corcos, L. A. Orozco, and S. Deffner, Phys. Rev. Lett. \textbf{114}, 233602
(2015).

\bibitem{Xu2} C. Liu, Z.-Y. Xu, and S. Zhu, Phys. Rev. A \textbf{91}, 022102
(2015).

\bibitem{speedup4} Y.-J. Zhang, W. Han, Y.-J. Xia, J.-P. Cao, and H. Fan,
Phys. Rev. A \textbf{91}, 032112 (2015).

\bibitem{speedup5} H.-B. Liu, W. L. Yang, J.-H. An, and Z.-Y. Xu, Phys. Rev.
A \textbf{93}, 020105(R) (2016).

\bibitem{speedup6} X. Cai and Y. Zheng, Phys. Rev. A \textbf{95}, 052104
(2017).

\bibitem{metrology1} V. Giovanetti, S. Lloyd, and L. Maccone, Nat. Photon.
\textbf{5}, 222 (2011).

\bibitem{metrology2} A. W. Chin, S. F. Huelga, and M. B. Plenio, Phys. Rev.
Lett. \textbf{109}, 233601 (2012).

\bibitem{QB} F. Campaioli, F. A. Pollock, F. C. Binder, L. C\'{e}leri, J.
Goold, S. Vinjanampathy, and K. Modi, Phys. Rev. Lett. \textbf{118}, 150601
(2017).

\bibitem{Rice} M. Demirplack and S. A. Rice, J. Chem. Phys. \textbf{129},
154111 (2008).

\bibitem{Berry} M. Berry, J. Phys. A \textbf{42}, 365303 (2009).

\bibitem{STA-review} E. Torrontegui, S. S. Ib\'{a}\~{n}ez,, S. Mart\'{\i}%
nez-Garaot, M. Modugno, A. del Campo, D. Gu\'{e}ry-Odelin, A. Ruschhaupt, X.
Chen, and J. G. Muga, Adv. At. Mol. Opt. Phys. \textbf{62}, 117 (2013).

\bibitem{STA10} X. Chen, I. Lizuain, A. Ruschhaupt, D. Gu\'{e}ry-Odelin, and
J. G. Muga, Phys. Rev. Lett. \textbf{105}, 123003 (2010).

\bibitem{STA11} J. G. Muga, X. Chen, S. Ib\'{a}\~{n}ez, I. Lizuain, and A.
Ruschhaupt, J. Phys. B \textbf{43}, 085509 (2010).

\bibitem{STA-exp1} M. G. Bason, M. Viteau, N. Malossi, P. Huillery, E.
Arimondo, D. Ciampini, R. Fazio, V. Giovannetti, R. Mannella, and O. Morsch,
Nat. Phys. \textbf{8}, 147 (2011).

\bibitem{STA21} A. del Campo, M. M. Rams, and W. H. Zurek, Phys. Rev. Lett.
\textbf{109}, 115703 (2012).

\bibitem{STA31} C. Jarzynski, Phys. Rev. A \textbf{88}, 040101(R) (2013).

\bibitem{shortcuts-NV} J. Zhang, J. H. Shim, I. Niemeyer, T. Taniguchi, T.
Teraji, H. Abe, S. Onoda, T. Yamamoto, T. Ohshima, J. Isoya, and D. Suter,
Phys. Rev. Lett. \textbf{110}, 240501 (2013).

\bibitem{STA32} A. del Campo, Phys. Rev. Lett. \textbf{111}, 100502 (2013).

\bibitem{STA40} S. Deffner, C. Jarzynski, and A. del Campo, Phys. Rev. X
\textbf{4}, 021013 (2014).

\bibitem{STA41} G. Vacanti, R. Fazio, S. Montangero, G. M. Palma, M.
Paternostro, and V. Vedral, New J. Phys. \textbf{16}, 053017 (2014).

\bibitem{santos} A. C. Santos and M. S. Sarandy, Sci. Rep. \textbf{5}, 15775
(2015).

\bibitem{STA50} S. Deffner, New J. Phys. \textbf{18}, 012001 (2015).

\bibitem{STA60} S. An, D. Lv, A. del Campo, and K. Kim, Nat. Commun. \textbf{%
7}, 12999 (2016).

\bibitem{santos2} A. C. Santos, R. D. Silva, and M. S. Sarandy, Phys. Rev. A
\textbf{93}, 012311 (2016).

\bibitem{song1} X.-K. Song, Q. Ai, J. Qiu, and F.-G. Deng, Phys. Rev. A
\textbf{93}, 052324 (2016).

\bibitem{cost} Y. Zheng, S. Campbell, G. De Chiara, and D. Poletti, Phys.
Rev. A \textbf{94}, 042132 (2016).

\bibitem{STA61} Y.-H. Chen, Y. Xia, Q.-C. Wu, B.-H. Huang, and J. Song,
Phys. Rev. A \textbf{93}, 052109 (2016).

\bibitem{NX} B. B. Zhou, A. Baksic, H. Ribeiro, C. G. Yale, F. J. Heremans,
P. C. Jerger, A. Auer, G. Burkard, A. A. Clerk, and D. D. Awschalom, Nat.
Phys. \textbf{13}, 330 (2017).

\bibitem{speedcost} S. Campbell and S. Deffner, Phys. Rev. Lett. \textbf{118}%
, 100601 (2017).

\bibitem{work} K. Funo, J.-N. Zhang, C. Chatou, K. Kim, M. Ueda, and A. del
Campo, Phys. Rev. Lett. \textbf{118}, 100602 (2017).

\bibitem{song2} X.-K. Song, F.-G. Deng, L. Lamata, and J. G. Muga, Phys.
Rev. A \textbf{95}, 022332 (2017).

\bibitem{STA70} E. Torrontegui, I. Lizuain, S. Gonz\'{a}lez-Resines, A.
Tobalina, A. Ruschhaupt, R. Kosloff, J. G. Muga, Phys. Rev. A \textbf{96},
022133 (2017).

\bibitem{book-quantum} J. J. Sakurai and J. J. Napolitano, \textit{Modern
Quantum Mechanics }(New York, Addison-Wesley, 2010).

\bibitem{Note} Note that the individual driving is equivalent to the
collective driving, i.e., $H^{A}(t)=H_{a}^{A}(t)=H_{b}^{A}(t)$, for a
quantum system with only two energy eigenstates \{$\left\vert
a_{t}\right\rangle ,\left\vert b_{t}\right\rangle $\}. The proof is
straightforward by using $\left\vert a_{t}\right\rangle \left\langle
a_{t}\right\vert +\left\vert b_{t}\right\rangle \left\langle
b_{t}\right\vert =1$.

\bibitem{Note2} The proof is straightforward with Eqs. (\ref{cost-relation})
and (\ref{condition}): $\partial _{t}C=[(1/2)\sum_{n}\left( \partial
_{t}C_{n}\right) ^{2/\alpha }]^{\alpha /2}=[(1/2)\left( \partial
_{t}C_{k}\right) ^{2/\alpha }+(1/2)\sum_{n\neq k}\left( \partial
_{t}C_{n}\right) ^{2/\alpha }]^{\alpha /2}=[(1/2)\left( \partial
_{t}C_{k}\right) ^{2/\alpha }+(1/2)\left( \partial _{t}C_{k}\right)
^{2/\alpha }]^{\alpha /2}=\partial _{t}C_{k}.$

\bibitem{book-dg} M. Fecko, \textit{Differential Geometry and Lie Groups for
Physicists}, (Cambridge University Press, 2006).

\bibitem{book-gs} I. Bengtsson and K. \.{Z}yczkowski, \textit{Geometry of
Quantum States: An Introduction of Entanglement}, (Cambridge University
Press, 2017).

\bibitem{Note3} There exist many monotone Riemannian metrics for mixed
states \cite{book-gs}. However, only quantum Fisher information metric can
exactly reproduce the Fubini--Study metric in pure state case \cite{book-gs}%
. In our paper, we will consider both canonical ensemble state (mixed state)
and single energy eigenstate (pure state) as the initial states, therefore,
the quantum Fisher information metric becomes the most appropriate choice.

\bibitem{fidelity} M. H\"{u}bner, Phys. Lett. A \textbf{163}, 239 (1992).

\bibitem{Caves} S. L. Braunstein and C. M. Caves, Phys. Rev. Lett. \textbf{72%
}, 3439 (1994).

\bibitem{Xu} Z.-Y. Xu, New J. Phys. \textbf{18}, 073005 (2016).

\bibitem{spin} In this paper, we adopt $S_{x}=\frac{\hbar }{\sqrt{2}}\left(
\begin{array}{ccc}
0 & 1 & 0 \\
1 & 0 & 1 \\
0 & 1 & 0%
\end{array}%
\right) $, $S_{y}=\frac{\hbar }{\sqrt{2}}\left(
\begin{array}{ccc}
0 & -i & 0 \\
i & 0 & -i \\
0 & i & 0%
\end{array}%
\right) $, and $S_{z}=\hbar \left(
\begin{array}{ccc}
1 & 0 & 0 \\
0 & 0 & 0 \\
0 & 0 & -1%
\end{array}%
\right) $ as the components of the spin-$1$ operator.

\bibitem{NV-Review1} M. W. Doherty, N. B. Manson, P. Delaneyc, F. Jelezko,
J. Wrachtrup, L. Hollenberg, Phys. Rep. \textbf{528}, 1 (2013).

\bibitem{NV-Review2} D. Suter and F. Jelezko, Prog. Nucl. Magn. Reson.
Spectrosc. \textbf{98-99}, 50 (2017).

\bibitem{LZS-NV} P. Huang, J. Zhou, F. Fang, X. Kong, X. Xu, C. Ju, and J.
Du, Phys. Rev. X \textbf{1}, 011003 (2011).

\bibitem{LZS-NV2} J. Zhou, P. Huang, Q. Zhang, Z. Wang, T. Tan, X. Xu, F.
Shi, X. Rong, S. Ashhab, and J. Du, Phys. Rev. Lett. \textbf{112}, 010503
(2014).

\bibitem{ion} T.\thinspace P. Xiong, L.\thinspace L. Yan, F. Zhou, K. Rehan,
D.\thinspace F. Liang, L. Chen, W.\thinspace L. Yang, Z.\thinspace H. Ma, M.
Feng, and V. Vedral, Phys. Rev. Lett. \textbf{120}, 010601 (2018).
\end{thebibliography}
\end{document}